\documentclass[preprint,prb,superscriptaddress,amssymb,amsmath, floatfix,citeautoscript]{revtex4-1}

\usepackage[pdftex]{graphicx}

\newcommand{\p}{\partial }
\newcommand{\w}{\omega}

\newcommand{\vw}{v_{\w}}

\newcommand{\lw}{\Lambda_{\w}}

\newcommand{\cw}{C_{\w}}

\newcommand{\sat}{AgSbTe$_{2}$}

\begin{document}

\title{Towards a microscopic understanding of phonon heat conduction}
%
\author{A. J. Minnich}
\email{aminnich@caltech.edu}
\affiliation{Division of Engineering and Applied Science, California Institute of Technology, Pasadena, CA 91125}
\maketitle    

Heat conduction by phonons is a ubiquitous process that incorporates a wide range of physics and plays an essential role in applications ranging from space power generation to LED lighting. Heat conduction has been studied for over two hundred years, yet many microscopic aspects of heat conduction have remained unclear in most crystalline solids, including which phonons carry heat and how natural and artificial structures scatter specific phonons. Fortunately, recent advances in both computation and experiment are enabling an unprecedented microscopic view of thermal transport by phonons. In this topical review, we provide an overview of these methods, the insights they are providing, and their impact on the science and engineering of heat conduction.

\clearpage

\section{Introduction}

Heat conduction by phonons in solids is a fundamental process that incorporates a variety of fascinating physics in addition to playing a central role in numerous applications. At the macroscale, heat conduction is well described by diffusion theory based on Fourier's law, which states that the heat flux is linearly related to a temperature gradient by a material property, the thermal conductivity. However, accessible length and time scales have dramatically shrunk over the past 30 years, and when these scales are comparable to phonon mean free paths (MFPs) and relaxation times, respectively, macroscopic theories do not provide an accurate picture of the transport \cite{Cahill:2014,Cahill:2003}. For example, Fourier's law substantially underpredicts the temperature rises of localized hotspots at the junctions of transistors \cite{Pop:2006,Pop:2010}. New nanostructured materials have been created with substantially reduced thermal conductivities compared to the bulk material, and some of these materials are under development as thermoelectric materials \cite{Biswas:2012,Mehta:2012,Voneshen:2013,Chowdhury:2009,Ma:2013b,SciencePaper}. The lifetimes of LEDs for domestic lighting as well as the performance of high power transistors is limited by near-junction thermal resistances due to defects and interfaces near the active region of these devices \cite{Yan:2012,Su:2012a,Cho:2014}.

To account for these effects, a microscopic picture of the heat conduction is necessary, including details such as which phonons among the broad thermal spectrum carry heat and how phonons interact with both natural and artificial structures. Unfortunately, this information has historically been very difficult to access. Consider the relation between thermal conductivity $k$ and microscopic quantities, given by the kinetic equation \cite{GangBook}:

\begin{equation}
k = \frac{1}{3} \int_{0}^{\w_{m}} \cw \vw \lw d\w
\end{equation}
where $\cw$ is the frequency-dependent specific heat, $\vw$ is the group velocity, $\lw$ is the MFP, and $\w$ is the phonon frequency. While the specific heat and group velocity are set by the harmonic component of the interatomic potential and can be measured using techniques such as neutron scattering, the MFPs are determined by deviations from the perfect harmonic lattice, either from the anharmonic component of the interatomic potential or from physical defects, and are very difficult to measure directly or calculate quantitatively. 

Despite these difficulties, considerable progress was made in building a microscopic picture of heat conduction using macroscopic measurements and solid state physics. Debye first attributed heat conduction to delocalized lattice waves and was able to explain the $1/T$ temperature dependence of thermal conductivity at sufficiently high temperatures \cite{Debye:1914}. Peierls then introduced the phonon Boltzmann equation and identified the two anharmonic interactions, normal and Umklapp processes, that return phonons to thermal equilibrium \cite{Peierls:1929}. Subsequently, approximate expressions for relaxation times due to different scattering mechanisms such as anharmonicity, point defects, and dislocations were derived by Pomeranchuk, Klemens, and Herring, among others. \cite{Herring:1954,Pomeranchuk:1941, KlemensSSP} In the early 1960s, Callaway, Holland, and others introduced models of thermal conductivity that could reasonably explain the temperature dependent thermal conductivity of pure crystals and alloys. \cite{Holland:1963,Callaway:1959, Steigmeier:1964} In fact, these models are still in frequent use today to study thermal transport in a wide range of materials. Microscopic phonon properties including the MFPs $\lw$ can be obtained by adjusting model parameters to fit macroscopic thermal conductivity data.

While this approach does yield useful insight, it has substantial limitations. First, the approach is not predictive because the fitting parameters must be determined using experimental data before any conclusions can be drawn. Second, the extracted MFPs strongly depend on assumptions made in the fitting and thus are difficult to determine unambiguously. For example, assuming a solid has a Debye dispersion leads to the contribution of high frequency phonons being substantially overpredicted because actual zone-edge phonons have a much smaller group velocity than the sound velocity \cite{Zebarjadi:2012,Jeong:2011,Zebarjadi:2011,AustinEES}. Further, if multiple scattering mechanisms with different temperature dependencies are present, as in complex thermoelectric materials, separating the various mechanisms is challenging. These ambiguities fundamentally occur because thermal conductivity represents an average over all phonon modes that results in a loss of the microscopic details of thermal transport.

Due to the lack of knowledge of MFPs and computational limits in solving the Boltzmann transport equation (BTE), for many years microscale heat conduction was studied with a grey model, which assumes that all phonons possess an average MFP. However, with this assumption one finds that there are many experimental observations that cannot be explained. Consider, for example, measurements by Song and Chen of the thermal conductivity of silicon films lithographically patterned with holes of approximately 2 micron diameter \cite{Song:2004}. If we take the average MFP in Si to be 300 nm at room temperature \cite{Ju:1999}, size effects would not be expected to play a role in the transport. However, substantially lower thermal conductivity than predicted by effective medium theory was observed in the actual sample. This result demonstrates that phonons in Si must have a broad spectrum and that some of the modes in this spectrum must possess MFPs on micron length scales. In fact, recent theoretical and experimental works have demonstrated that nearly half of the heat in silicon is contributed by these long MFP phonons \cite{Henry:2008, Regner:2013, Esfarjani:2011}. This example demonstrates the importance of having an accurate microscopic understanding of heat conduction to predict even the simplest phenomena at micron length scales.

Fortunately, a number of advances in both computation and experiment are providing a detailed microscopic picture of heat conduction in solids. In computation, ab initio and mesoscale techniques are enabling a multiscale understanding of phonons in pure and nanostructured crystals, often without any adjustable parameters. In experiment, improvements in inelastic neutron scattering techniques, as well as the use of a new experimental technique called mean free path spectroscopy, are enabling the first direct measurements of the transport properties resolved across the thermal phonon spectrum. In this topical review, we provide an overview of the recent insights into heat conduction and their impact on the science and engineering of thermal conductivity.

We note that a number of excellent reviews in the general area of nanoscale thermal transport have recently been published. \cite{Cahill:2014, Tian:2013, Luo:2013a, Zebarjadi:2011, Feng:2014} In this review, we focus specifically on advances in computation and experiment for studying the microscopic properties of thermal phonons responsible for heat conduction in crystals. The reader is referred to other reviews for a broader view of the nanoscale thermal transport field.

\section{Computation}

A number of advances have been made in computational techniques ranging from ab initio calculations using density functional theory (DFT) to mesoscale simulation with the Boltzmann transport equation (BTE). When employed together, these techniques are enabling the first multiscale simulations of thermal transport without any adjustable parameters. Here, we highlight recent results obtained from ab initio calculations, atomistic Green's functions, and variance-reduced MC methods to solve the BTE.

\subsection{Ab Initio}

Ab initio calculations of thermal conductivity in pure crystals are based on using DFT to determine the interatomic potential of the atoms in a crystal, from which thermal properties can be obtained using lattice dynamics or molecular dynamics. Prior to this approach, atomistic calculations were based on semi-empirical potentials such as the Stillinger-Weber potential with adjustable parameters for each material. However, these potentials were designed to fit experimentally measured lattice constants and elastic constants, and thermal conductivity predictions were often in poor agreement with the actual values \cite{Broido:2005}. Further, empirical potentials have limited predictive power because the fitting parameters must be determined by fitting to experimental data. In contrast, the ab initio approach is able to accurately calculate the interatomic potential without any adjustable parameters or any other inputs other than fundamental constants. As a result, this predictive approach is providing a new understanding of the origin of a material's thermal conductivity.

DFT has long been used to determine phonon dispersions by calculating the harmonic force constants between atoms\cite{Yin:1982, Giannozzi:1991, Baroni:2001, Baroni:1987}. Yin and Cohen reported calculations of nonlinear response coefficients such as cubic force constants at high symmetry points by taking finite differences of harmonic force constants. Subsequent work showed that the anharmonic force constants could be obtained using density functional perturbation theory (DFPT) and the $2n+1$ theorem for certain crystals \cite{Gonze:1989}. This approach was used to calculate the linewidths of high symmetry points in the Brillouin zone, enabling comparison to Raman measurements with excellent agreement \cite{Debernardi:1995, Debernardi:1998}. Deinzer et al. first reported the calculation of linewidths over the entire Brillouin zone using DFPT. \cite{Deinzer:2003} Broido et al. then extended this approach to compute the thermal conductivities of Si and Ge without any adjustable parameters using an exact iterative solution of the Boltzmann transport equation \cite{Broido:2007}. Later works used a real-space approach based on calculating forces due to systematic atomic displacements using DFT that can be applied to complicated crystal structures \cite{Esfarjani:2008}. These methods have now been applied to a wide variety of materials with resounding success, ranging from pure and compound semiconductors \cite{Ward:2010, Li:2012h, Broido:2013, Lindsay:2013,  Li:2013e, Esfarjani:2011, Luo:2013, Liao:2014, Wee:2010, Ward:2009a}, isotopically impure crystals \cite{Lindsay:2013b, Mingo:2010, Lindsay:2013a,Lindsay:2012a, Stewart:2009}, nanowires \cite{Li:2012e}, nanotubes \cite{Mingo:2008,Lindsay:2009},materials under extreme pressure \cite{Broido:2012}, half Heuslers \cite{Shiomi:2011}, PbTe \cite{Shiga:2012, Tian:2012, Murakami:2013}, Bi \cite{Lee:2014a}, lead chalcogenides \cite{lee_resonant_2014}, alloys \cite{Garg:2011a}, superlattices \cite{Garg:2013, Garg:2011}, to 2D materials such as MoS$_{2}$ \cite{Li:2013g} and graphene \cite{Bonini:2012a, Paulatto:2013}.

It is not an exaggeration to say that these calculations have fundamentally changed our understanding of microscopic origin of thermal conductivity. We refer the reader to a recently published review for more information on the computational details of the ab initio approach \cite{Feng:2014}. Here, we instead highlight a small sampling of the insights gained into heat conduction. 

\begin{figure}
\begin{center}
\includegraphics[width=.9\textwidth]{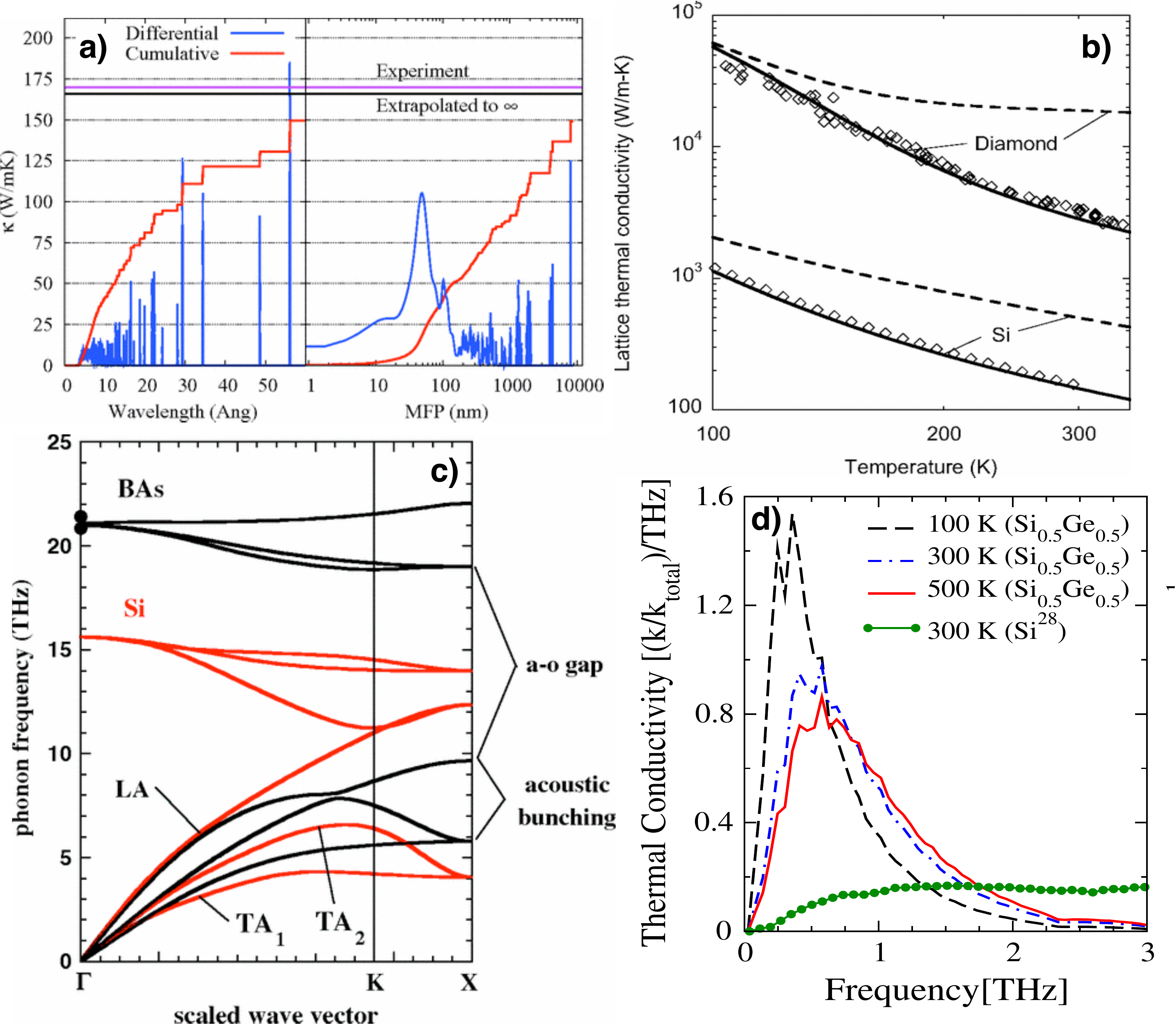}
\caption{(a) Spectral thermal conductivity of silicon at 277 K versus wavelength (left) and MFP (right). Nearly half the heat is carried by phonons with MFPs exceeding 1 micron, despite their small contribution to heat capacity. Reprinted Fig.\ 6 with permission from Ref.\ \onlinecite{Esfarjani:2011}. Copyright (2011) by the American Physical Society. (b) Thermal conductivity of diamond and silicon versus temperature with (solid line) and without (dashed line) optical phonon scattering. Optical phonons play a key role in setting thermal conductivity by providing channels for phonon-phonon scattering, although they do not carry substantial heat in bulk materials. Reprinted Fig.\ 5 with permission from Ref.\ \onlinecite{Ward:2009a}. Copyright (2009) by the American Physical Society. (c) Microscopic mechanisms underlying the predicted high thermal conductivity of BAs, including acoustic phonon bunching and a large acoustic-optical phonon gap that inhibit phonon-phonon scattering. Reprinted Fig.\ 5 with permission from Ref.\ \onlinecite{Lindsay:2013a}. Copyright (2013) by the American Physical Society. (d) Spectral thermal conductivity versus phonon frequency for Si$_{0.5}$Ge$_{0.5}$, demonstrating that the thermal conductivity is primarily due to phonons with frequencies less than 1 THz, in contrast to bulk Si. Reprinted Fig.\ 2 with permission from Ref.\ \onlinecite{Garg:2011a}. Copyright (2011) by the American Physical Society.}
\label{fig:DFT}
\end{center}
\end{figure}

One particularly important realization is the breadth of the thermal phonon spectrum and correspondingly, the importance of low frequency, long MFP phonons to heat conduction. Due to complications in determining the MFPs of phonons as discussed in the introduction, many works treated the phonon spectrum using average properties in a grey approximation. Because simple estimates of the average phonon MFP in Si based on kinetic theory predict a MFP of at most a few hundred nanometers, low frequency phonons with frequency less than 1 THz would not be expected to contribute to heat conduction due to their small heat capacity. However, this estimation turns out to be very misleading. In fact, thermal phonons possess an extremely broad spectrum, with MFPs ranging from a few nanometers to 10 microns in Si at room temperature \cite{Esfarjani:2011}. A calculation of the accumulated thermal conductivity contribution versus wavelength and MFP by Esfarjani et al. \cite{Esfarjani:2011} is shown in Fig.\ \ref{fig:DFT}a, demonstrating that while most thermal phonons have similar wavelengths, MFPs vary by orders of magnitude. Further, phonons with MFPs longer than 1 micron contribute over 40\% of the total thermal conductivity. This result is consistent with earlier MD simulations by Henry and Chen \cite{Henry:2008} and demonstrates the importance of considering the phonon MFP spectrum, rather than an average MFP, for interpreting thermal measurements in bulk and nanostructured materials.

Another important insight is the importance of optical modes to heat conduction, although by an indirect route. Due to their small group velocity, the contribution of optical modes to heat conduction is typically neglected. Ward and Broido found this assumption to be a good one for bulk Si and Ge,\cite{Ward:2010} with optical modes contributing less than 10\% although this contribution can be larger in nanostructures \cite{Tian:2011} and in some bulk materials such as PbTe \cite{Tian:2012}. However, while optical modes typically do not contribute to heat conduction substantially, they provide extremely important scattering channels for acoustic phonons. Removal of these optical modes from the scattering channels completely changes the temperature dependence of the relaxation times and results in a factor of three increase in the thermal conductivity in solids like diamond and Si as shown in Fig.\ \ref{fig:DFT}b. Optical modes thus play a key role in three-phonon scattering processes and hence the thermal conductivity of crystals.

Yet another insight is a microscopic understanding of what makes a material a good thermal conductor. The traditional criteria for high thermal conductivity, light atomic masses and stiff harmonic bonds, are well known \cite{Slack:1973}. While these macroscopic criteria are useful, the small number of high thermal conductivity crystals such as silicon and diamond is already known and new materials have not been identified in some time. Using first-principles calculations, Lindsay et al. recently identified a new set of microscopic criteria for thermal conductors based on an analysis of the compound BAs that offers substantially more insight into the origin of high thermal conductivity \cite{Lindsay:2013a}. As illustrated in Fig.\ \ref{fig:DFT}c, these requirements include acoustic phonon bunching so that interactions among acoustic modes are restricted, a large acoustic-optical phonon energy gap so that optical phonons cannot participate in phonon-phonon scattering, and isotopically pure elements to eliminate point defect scattering. The authors report that BAs incorporates all of these mechanisms and thus could have a high thermal conductivity comparable to that of diamond. These requirements are not obvious from a macroscopic perspective, demonstrating the importance of the microscopic view.

The calculations are not limited to just pure and compound materials. Alloys have also been studied using DFT using the virtual crystal approximation by Garg et al.\cite{Garg:2011a} A subtle aspect of alloys is that they consist of randomly placed mass defects and thus are not periodic. Attempting to calculate properties by progressively increasing the size of a DFT supercell with random masses can yield erroneous results such as the group velocity tending to zero. Instead, one must employ the virtual crystal model and incorporate defect scattering as an additional scattering mechanism. Garg et al. used this approach to examine the thermal conductivity of Si-Ge alloys. The authors found that the point defect scattering mechanism dramatically increases the relative contribution of low frequency modes to heat conduction, as shown in Fig.\ \ref{fig:DFT}d, with the majority of the heat carried by sub-THz frequency phonons with MFPs between 200 nm and 3 microns. Unexpectedly, the authors also found that scattering of low frequency phonons is increased in alloys compared to the pure crystal due to a change in the vibrational eigenmodes of the random structure.

These highlights are just a few of the many results that have been provided by ab initio calculations. Other insights include a better understanding of normal processes \cite{Ward:2009a} and the effect of superlattice periodicity on thermal conductivity \cite{Garg:2011,Garg:2013}, among others. The ab initio approach has had a very important impact on our understanding of thermal conductivity, and further progress is expected as it is applied to a wider range of solids.

\subsection{Atomistic Green's functions}

While DFT provides the most detailed microscopic information, the approach is very computationally intensive and cannot be applied to domains larger than a few atomic unit cells. However, many structures such as grain boundaries and nanoparticles are too large for DFT yet sufficiently small that the wave nature of the phonon on the discrete atomic lattice must be considered. For these situations, the atomistic Green's functions (AGF) approach has emerged as a powerful technique to exactly calculate the interaction of phonons with extended structures assuming a harmonic lattice \cite{Zhang:2007}. We note that molecular dynamics is also a useful approach that incorporates anharmonicity but is restricted to high temperatures \cite{Chalopin:2012,McGaughey:2006,Landry:2009,Zuckerman:2008,Schelling:2002a}. Further information on this method is available in another review \cite{Feng:2014}.

Green's functions are widely used to solve differential equations. The phonon AGF method has its origins in nanoscale electron transport calculations, where it is known as the Non-Equilibrium Green's Function (NEGF) formalism \cite{Datta:2005,Datta:1997}. In this approach, the Schrodinger equation is solved in a specified domain between two contacts, yielding the transmission function across the domain as a function of energy. The advantage of this technique is that interactions with semi-infinite contacts are exactly accounted for through self-energy terms. NEGF has been extensively used to simulate nanoscale electronic devices \cite{Datta:2000}.

 By making a few careful substitutions, an analogous approach can be developed for phonons \cite{Zhang:2007, Mingo:2009a}. This adaptation was introduced by Mingo and Yang to study dielectric nanowires coated with an amorphous material \cite{Mingo:2003}. Subsequently, AGF has been used to study phonon transport in a wide variety of structures, including Si/Ge interfaces \cite{Zhang:2007a}, nanowire junctions \cite{Zhang:2007b}, BN nanotubes \cite{Stewart:2009}, carbon nanotube pellets \cite{Chalopin:2009}, solids containing nanoparticles \cite{Kundu:2011}, and others. \cite{Pernot:2010,Huang:2010,Huang:2010a,Li:2012j,Li:2012g, Luckyanova:2012, Hopkins:2009e,Hopkins:2009d}

\begin{figure}
\begin{center}
\includegraphics[width=1\textwidth]{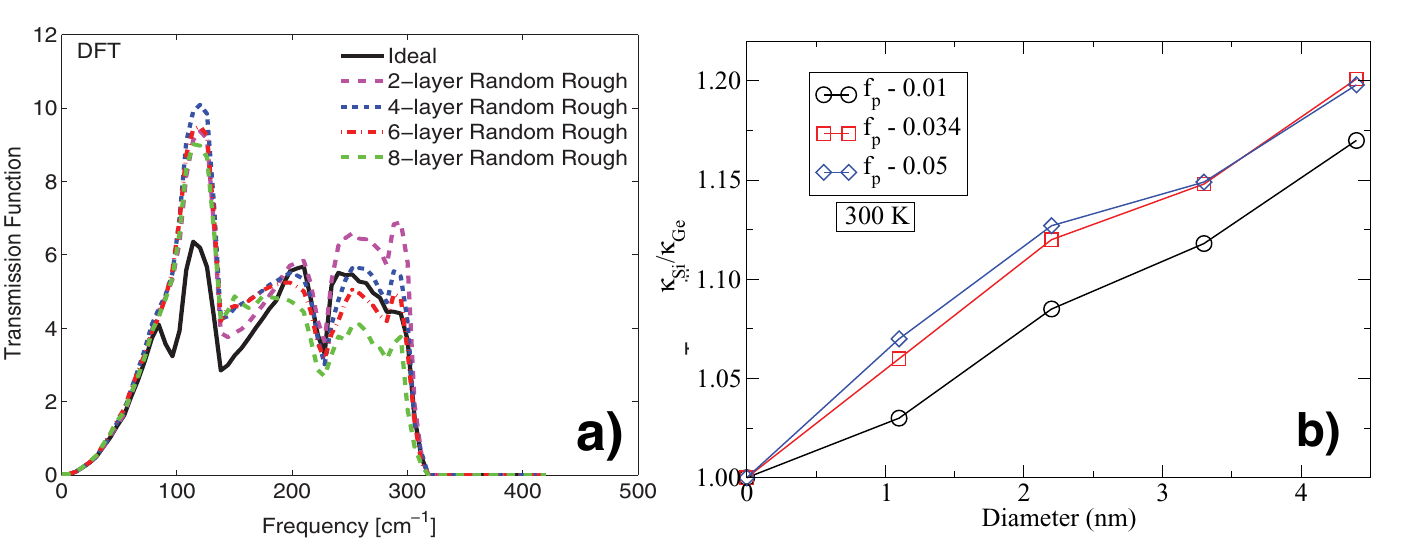}
\caption{(a) Transmittance from Si to Ge calculated using AGF with DFT force constants. Interfaces with atomic mixing, denoted ``rough'' interfaces, can enhance transmission across an interface. Reprinted Fig.\ 2 with permission from Ref.\ \onlinecite{Tian:2012a}. Copyright (2012) by the American Physical Society. (b) Ratio of thermal conductivity of a Si$_{0.5}$Ge$_{0.5}$ alloy with Si nanoparticles to that with Ge particles versus nanoparticle diameter. Several different nanoparticle concentrations $f_{p}$ are shown. Lighter nanoparticles composed of Si yield a higher thermal conductivity than heavier Ge nanoparticles of the same size. Reprinted Fig.\ 2 with permission from Ref.\ \onlinecite{Kundu:2011}. Copyright (2011) by the American Physical Society.}
\label{fig:AGF}
\end{center}
\end{figure}

We highlight two of the many recent works that have studied the interaction of phonons with extended structures. Tian et al. used AGF to study phonon transmission across ideal and rough heterogeneous Si/Ge interfaces using both ab initio and empirical force fields \cite{Tian:2012a}. The roughness was incorporated by mixing atoms near the Si/Ge interface. The authors found that the transmission of mid-range phonon frequencies is increased in rough interfaces compared to clean interfaces because atomic mixing softens the impedance mismatch between Si and Ge. Figure \ref{fig:AGF}a shows the calculated transmittance from Si to Ge, demonstrating that interfaces with a small amount of interfacial mixing can increase the interfacial transmittance and hence reduce the thermal boundary resistance. The authors also found the use of ab initio force fields critical to obtaining quantitative results.

Another work by Kundu et al. focused on calculating the scattering rate due to nanoparticles with diameter of a few nanometers without any adjustable parameters \cite{Kundu:2011}. Most treatments of nanoparticle scattering rely on continuum level theories that interpolate between Rayleigh and geometrical scattering limits \cite{Majumdar:1993}. Kundu et al. used AGF with ab initio force fields to exactly calculate the scattering rate from nanoparticles to all orders. They found that nanoparticles composed of heavier atoms than those of the host lattice can provide a larger reduction in thermal conductivity, as shown in Fig.\ \ref{fig:AGF}b. This result cannot be predicted from the Born approximation. The authors also found that simpler scattering rates based on the Born approximation and a geometrical scattering rate are reasonably accurate compared to the exact calculation, providing rigorous justification for the use of these simpler models.

\subsection{Variance-reduced Monte Carlo algorithms} \label{sec:MC}

Atomistic approaches have an advantage in that they require no adjustable parameters, but they are very computationally costly and cannot be applied to domains larger than a few nanometers. Simulating thermal transport in larger structures therefore requires a mesoscale treatment based on the Boltzmann transport equation (BTE). This equation has a long history and has been applied to model diverse phenomena ranging from photon transport through scattering atmospheres \cite{Chandrasekhar:1950} to neutron transport in nuclear reactors \cite{Sykes:1958}. The equation was introduced for phonons in solids including the complete anharmonic collision term by Peierls in 1929, \cite{Peierls:1929} and solutions of the BTE with simpler collision terms were obtained using mathematical techniques used in neutron transport theory by Engleman as early as 1958 \cite{Englman:1958}.


The BTE under the single-mode relaxation time approximation is given by:

\begin{equation} \label{}
\frac{\p f}{\p t} + \mathbf{v} \cdot \nabla f = - \frac{f - f_{0}}{\tau}
\end{equation}

where $f$ is the desired distribution function, $\mathbf{v}$ is the group velocity, $f_{0}$ is the equilibrium distribution that is related to $f$ by an integral equation, and $\tau$ is the relaxation time that depends on phonon frequency. In general, this integro-differential equation is a function of 8 variables - time, phonon frequency, three spatial variables, and three momentum space variables, making its solution very challenging.

In certain situations, exact analytical solutions can be obtained using integral transforms in infinite or semi-infinite domains \cite{Placzek:1947,Mark:1947}. We have recently reported a solution for an infinite domain including frequency dependence \cite{Hua:2014}, and Collins et al.\ reported solutions for different spectral models \cite{Collins:2013}. However, in most realistic situations, a numerical solution is required. The most straightforward numerical strategy is simply to apply finite differences to the derivatives, discretize the integrals, and solve the equation in space and time, an approach known as discrete ordinates \cite{Chandrasekhar:1950}. This approach for phonons was reported by Majumdar in a 1D geometry \cite{Majumdar:1993, Joshi:1993}. Recently, discrete ordinates has been used to simulate steady 2D transport under the grey approximation \cite{Yang:2005a} and 1D transient transport including frequency dependence \cite{Minnich:2011}. However, discrete ordinates is challenging to apply to multiple spatial dimensions while including frequency dependence due to the substantial memory requirements. Other approaches taken in the thermal transport field include a two-flux method \cite{Rowlette:2008,Sinha:2006a,Sinha:2006}, Monte Carlo \cite{Klitsner:1988, Peterson:1994,Lacroix:2006,Mazumder:2001,Mittal:2010,Hao:2009,Jeng:2008}, finite volume \cite{Narumanchi:2003,Narumanchi:2005,Narumanchi:2004}, a mean free path sampling algorithm \cite{McGaughey:2012}, and a lattice Boltzmann solver \cite{Heino:2010}. However, these algorithms are either computationally costly or make simplifying approximations that limit the validity of the solution.

\begin{figure}
\begin{center}
\includegraphics[width=0.8\textwidth]{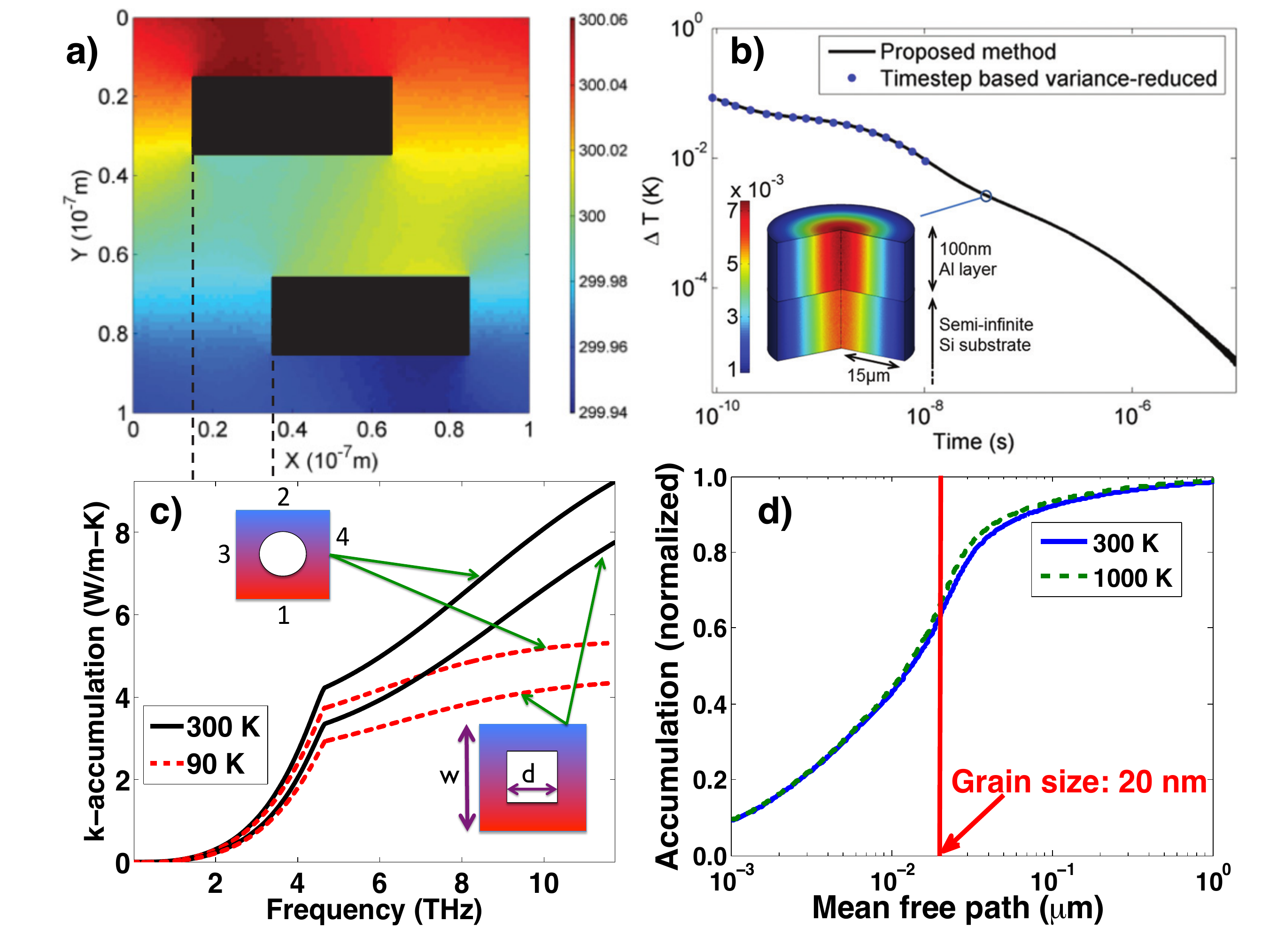}
\caption{(a) Temperature field of silicon with displaced rectangular pores subject to a small temperature gradient, demonstrating extremely low stochastic noise.  Reprinted Fig.\ 10 with permission from Ref.\ \onlinecite{Peraud:2011}. Copyright (2011) by the American Physical Society. (b) Calculation of surface temperature versus time in the geometry of a TDTR experiment, consisting of a film on a substrate with a Gaussian radial initial temperature distribution in the film. The more efficient linearized algorithm \cite{Peraud:2012} can calculate the temperature decay to microsecond time scales with little stochastic noise, compared to nanosecond time scales possible with the original algorithm. Reprinted with permission from Ref.\ \onlinecite{Peraud:2012}. Copyright (2012), AIP Publishing LLC. (c) Spectral thermal conductivity accumulation versus phonon frequency for a nanomesh, or a thin silicon membrane with periodic pores. The heat carried by low frequency phonons that could undergo coherent interference is not sufficient to explain the exceptionally low thermal conductivities of 1-2 W/mK reported in nanomeshes. \cite{Ravichandran:2014} (d) MFP accumulation function for nanocrystalline silicon with a 550 nm grain size including a frequency-dependent grain boundary scattering rate. Due to the frequency-dependent scattering, long MFP phonons still contribute substantially to heat conduction despite the presence of grain boundaries. \cite{Hua:2014a} }
\label{fig:MC}
\end{center}
\end{figure}

Recently, variance-reduced Monte Carlo (MC) algorithms have been introduced that solve the BTE many orders of magnitude faster than other algorithms in complex 3D geometries while rigorously including frequency-dependence \cite{Peraud:2012,Radtke:2009,Peraud:2011,Hadjiconstantinou:2010}. These variance-reduced algorithms were originally introduced to simulate rarified gas dynamics \cite{Homolle:2007,Baker:2005} and have been adapted for phonons by Radtke, Peraud and Hadjiconstantinou \cite{Peraud:2012,Radtke:2009,Peraud:2011}. The particular variance-reduced algorithm adapted for phonons is based on a statistical method called control variates in which the incorporation of deterministic information can reduce the variance of an estimator. For phonons, the deterministic knowledge is the fact that the equilibrium Bose-Einstein distribution is known analytically. Rather than simulating this known distribution stochastically, in deviational MC only the deviation from this equilibrium distribution is computed, thereby dramatically reducing the variance. An example of the low stochastic noise result that can be obtained is shown in Fig.\ \ref{fig:MC}a, showing the two dimensional temperature profile of a slab with square pores calculated by Peraud et al. \cite{Peraud:2011}

This algorithm is most efficient compared to traditional MC when the maximum temperature variation over the domain is much smaller than the equilibrium temperature \cite{Hadjiconstantinou:2010}, with the efficiency enhancement improving as $(\Delta T/T)^{2}$. In the special case when $\Delta T/T \ll 1$, an additional simplification can be made that reduces the computational cost by an additional few orders of magnitude \cite{Peraud:2012}. In this situation, the distribution to which scattering phonons relax is nearly the same everywhere in the domain due to the small temperature differential, allowing the scattering operator to be linearized. This simplification allows the particles to be simulated completely independently from each other and without any need for temporal and spatial discretization, simplifying parallelization and reducing memory requirements by orders of magnitude. 

These algorithms, and the latter linearized algorithm in particular, are enabling new insights into mesoscale thermal transport that have not been previously possible due to computational limitations. For example, Peraud and Hadjiconstantinou demonstrated the efficiency of their algorithm by performing the first fully 3D simulations of a common optical experiment, time-domain thermoreflectance, to microsecond time scales, as shown in Fig.\ \ref{fig:MC}b \cite{Peraud:2012}. We have also implemented this simulation to better understand recent observations of quasiballistic transport using variable pump-size TDTR measurements \cite{Ding:2014}, as will be described in section \ref{MFPS}.

In our group, we are using these algorithms to study a number of problems that were not possible to address previously due to computational limitations. One problem we have examined is the origin of exceptionally low thermal conductivities in silicon nanomeshes, consisting of periodic holes etched in a thin membrane \cite{Yu:2010, Tang:2010}. The measured thermal conductivities were apparently too low to be explained by boundary scattering and thus tentatively attributed to coherent phonon interference. Other works attempted to simulate phonon transport in these structures using the BTE but could only incorporate two dimensions \cite{Hao:2009} or modeled the scattering with a phenomenological scattering rate that lacked predictive power \cite{Dechaumphai:2012}. 

Using variance-reduced MC algorithms, we examined the origin of these observations by simulating the full 3D geometry of the nanomeshes while rigorously including the frequency-dependence of phonon properties \cite{Ravichandran:2014}. As shown in inset of Fig.\ \ref{fig:MC}c, we simulated thermal transport through one unit cell of a nanomesh with circular and square pores using periodic heat flux boundary conditions \cite{Hao:2009} and diffuse boundary scattering from pore walls and top and bottom boundaries. Our calculation demonstrated that the heat carried by low frequency modes that were most likely to undergo coherent interference was not sufficient to explain the observed low thermal conductivies of 1-2 W/mK. As shown in Fig.\ \ref{fig:MC}c, most of the heat in nanomeshes is carried by phonons with frequencies larger than 3 THz, corresponding to wavelengths of less than two nanometers. These wavelengths are very small compared to the reported periodicity of 10-20 nanometers, thus making coherent effects unlikely to affect heat conduction. Instead, the low thermal conductivities could be accounted for by the presence of a 2-3 nm thick native oxide that effectively increased the pore size as well as phonon backscattering at the pore walls. Our calculations demonstrate that thermal phononic crystals at room temperature must have periodicity on the order of 1-2 nanometers and atomic level roughness, a difficult fabrication requirement at present. However, as recently reported by Zen et al, thermal phononic effects can occur at sub-K temperatures even with micron-sized structures because of the increase in thermal phonon wavelength at these low temperatures \cite{zen_engineering_2014}.

We have also used these algorithms to study thermal transport in nanocrystalline silicon germanium alloys with a realistic 3D grain structure \cite{Hua:2014a}. A recent experimental work by Wang et al. provided evidence that the scattering rate due to grain boundary scattering in nanocrystalline silicon is not grey, as typically assumed, but must depend on the phonon frequency \cite{Wang:2011a}. We wanted to examine the impact of the frequency-dependent grain boundary scattering on the distribution of heat in the thermal phonon spectrum. While a phenomenological scattering rate $\tau^{-1}=v/L$ can be used along with an analytic solution of the BTE to obtain some insight, this approach lacks predictive power because in the non-grey model the geometrical length $L$ depends on phonon frequency and is not known in advance. Instead, we modeled phonon transmission across the grain boundary and exactly incorporated the geometrical effects using the linearized MC algorithm with a cubic, 3D grain structure and periodic heat flux boundary conditions \cite{Hao:2009}. We found that the scattering rate required to explain the measurements of Wang et al. resulted in a large fraction of heat carried by low frequency phonons with MFPs exceeding the grain size. Figure \ref{fig:MC}d shows the accumulated thermal conductivity versus phonon MFP, demonstrating the large contribution from long MFP phonons despite the presence of the grain boundary. Our results suggest that long MFP phonons may still contribute substantially to heat conduction in nanocrystalline materials, with important implications for improving the efficiency of thermoelectrics.

Many other interesting studies lie ahead using these powerful algorithms. When combined with the previously described first-principles and atomistic level modeling techniques, these algorithms enable simulations of thermal transport in realistic, multidimensional structures without any adjustable parameters.

\section{Experiment}

Experimentally measuring the transport properties of specific phonons has been a considerable challenge until recently. Here, we describe two experimental techniques, inelastic neutron scattering (INS) and mean free path (MFP) spectroscopy, that are providing a detailed, microscopic view of the thermal phonon spectrum.

\subsection{Inelastic neutron scattering}
INS has been used for decades in condensed matter physics to measure phonon dispersions by observing the interaction of a neutron with a phonon \cite{Cochran:1966, Alperin:1972}. Recently, INS has emerged as a powerful technique to provide microscopic measurements of the transport properties of phonons \cite{Delaire:2011,Ma:2013b}. By leveraging a number of technical advances, INS can now be used to map phonon modes and relaxation times over the entire Brillouin zone in single crystal samples.

In early INS implementations, continuous neutron fluxes and triple-axis spectrometers were used to serially map the scattering function versus wavevector and energy. This approach requires substantial beam time, restricting studies to high symmetry lines in the crystal. Recent advances in pulsed spallation sources, time-of-flight neutron spectrometers, and large-area detectors now allow a wide range of wavevectors to be measured simultaneously, dramatically reducing the beam time needed for measurements. Further, advances in software allow the full 4D scattering function $S(\mathbf{q}, E)$ to be reconstructed from several measurements with different crystal orientations. Finally, a better ability to correct for instrumental broadening and multiple scattering effects enables the phonon linewidths, and therefore relaxation times, to be accurately measured.

\begin{figure}
\begin{center}
\includegraphics[width=1\textwidth]{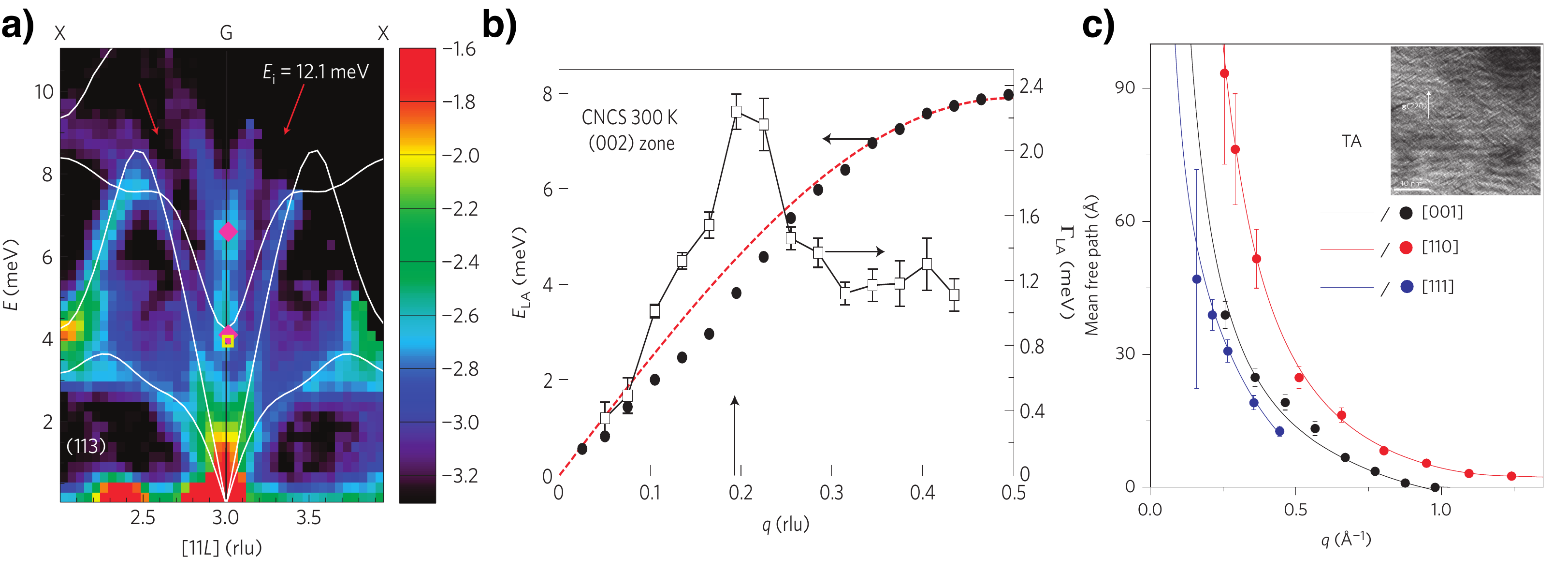}
\caption{(a) Phonon dispersion measurements for PbTe at 300 K using the time-of-flight Cold Neutron Chopper Spectrometer, demonstrating the existence of an avoided crossing between the LA and TO modes. Reprinted by permission from Macmillan Publishers Ltd: Nature Materials, Ref.\ \onlinecite{Delaire:2011}, Copyright (2011). (b) Measured dispersion and linewidths of the LA mode in PbTe \cite{Delaire:2011}, demonstrating a peak in the linewidth corresponding to a large scattering rate. Reprinted by permission from Macmillan Publishers Ltd: Nature Materials, Ref.\ \onlinecite{Delaire:2011}, Copyright (2011). (c) Measured TA phonon MFPs of \sat \ along different crystallographic directions. The MFPs are remarkably short and of the order of the size of the nanodomains, shown in the TEM image in the inset. Reprinted by permission from Macmillan Publishers Ltd: Nature Nanotechnology, Ref.\ \onlinecite{Ma:2013b}, Copyright (2013). }
\label{fig:INS}
\end{center}
\end{figure}

These advances were first applied to PbTe, historically an excellent thermoelectric material with very low lattice thermal conductivity  \cite{Delaire:2011}. However, the microscopic origin of this low thermal conductivity remained unclear. Using INS, Delaire et al. mapped the full Brillouin zone of PbTe and uncovered an interesting feature of the phonon dispersion. While harmonic DFT calculations predict the LA and TO modes to cross, INS measurements revealed the presence of an extended avoided crossing as shown in Fig. \ref{fig:INS}a, indicating a strong anharmonic repulsion between the branches. This strong scattering was also reflected in the phonon linewidths, which exhibited a peak in the LA branch at certain frequencies due to the interaction as in Fig.\ \ref{fig:INS}b. The INS measurements thus demonstrated that low thermal conductivity of PbTe can be attributed in large part to the extremely strong damping of the LA branch by the LA-TO interaction. It is interesting to note that previous DFT calculations and INS measurements were unable to identify this avoided crossing, demonstrating the unique results provided by advances in the INS technique.

Subsequently, INS was used by Ma et al. to study a particularly puzzling material, \sat \cite{Ma:2013b, Minnich:2013}. The unusual thermal conductivity of this material was originally reported by Morelli et al. \cite{Morelli:2008} and is puzzling because it is clearly a crystal, as evidenced by x-ray diffraction, yet its thermal conductivity follows the trend of a glass rather than that a crystal, with a slow increase of thermal conductivity with temperature. Morelli et al. originally attributed this observation to extreme anharmonicity from the repulsion of a lone-electron pair.

Ma et al. applied a number of experimental techniques to this material, including INS, to provide a detailed microscopic account of the thermal conductivity of \sat. As was performed for PbTe, the authors mapped the scattering function and linewidths over the full Brillouin zone and again found interesting features. In \sat, the linewidths are extremely broad, even in comparison to PbTe, with the upper part of the LA and optical branches so broad that they merge into a continuum. However, no temperature dependence of linewidths was observed, demonstrating that anharmonicity could not be responsible for the low thermal conductivity as it was for PbTe. Because the scattering rate due to point defects was also calculated to be too weak, the INS measurements showed that another defect structure must be responsible.

The authors used TEM to confirm the presence of this defect structure in the form of ordered nanodomains that spontaneously form over length scales of a few nanometers, values that are consistent with the phonon MFPs determined from the measured linewidths as shown in Fig.\ \ref{fig:INS}c. DFT calculations showed that the reason for this spontaneous nanostructure is a degeneracy in ground state crystal structure energy. It is these nanodomains that are responsible for the low thermal conductivity, scattering phonons so strongly that the thermal conductivity approaches the amorphous limit. The authors noted that this spontaneous nanostructure may be useful for thermoelectrics because it is thermodynamically stable, unlike artificial nanostructures that are often metastable.

These two studies illustrate the utility of INS to provide an experimental microscopic view of thermal transport. Not only are the results of fundamental scientific interest, they also guide the development of more efficient thermoelectric materials.

\subsection{Mean free path spectroscopy} \label{MFPS}

INS, while a powerful technique, does have limitations. It is best suited for single crystals samples for which the full Brillouin zone can be mapped. Additionally, low energy phonons, which have recently been demonstrate to play an essential role in heat conduction \cite{Regner:2013, Henry:2008, Esfarjani:2011}, are difficult to study with INS due interference from the elastic scattering peak. 

Fortunately, a considerably simpler experimental technique has emerged in the past several years that enables the first direct MFP measurements over a wide range of length scales and materials using readily available equipment. The technique, mean free path spectroscopy \cite{TCPRL}, is able to directly measure the MFP accumulation function, defined as the accumulated thermal conductivity as a function of phonon MFP \cite{damesCRC, Yang:2013a}. This distribution contains less information than INS measurements on single crystals but still provides extremely useful insights into which phonons carry heat as well as the key length scales at which size effects occur. Further, the technique can be applied to most solids, including complex materials such as polycrystals or nanocomposites.

\begin{figure}
\begin{center}
\includegraphics[width=1\textwidth]{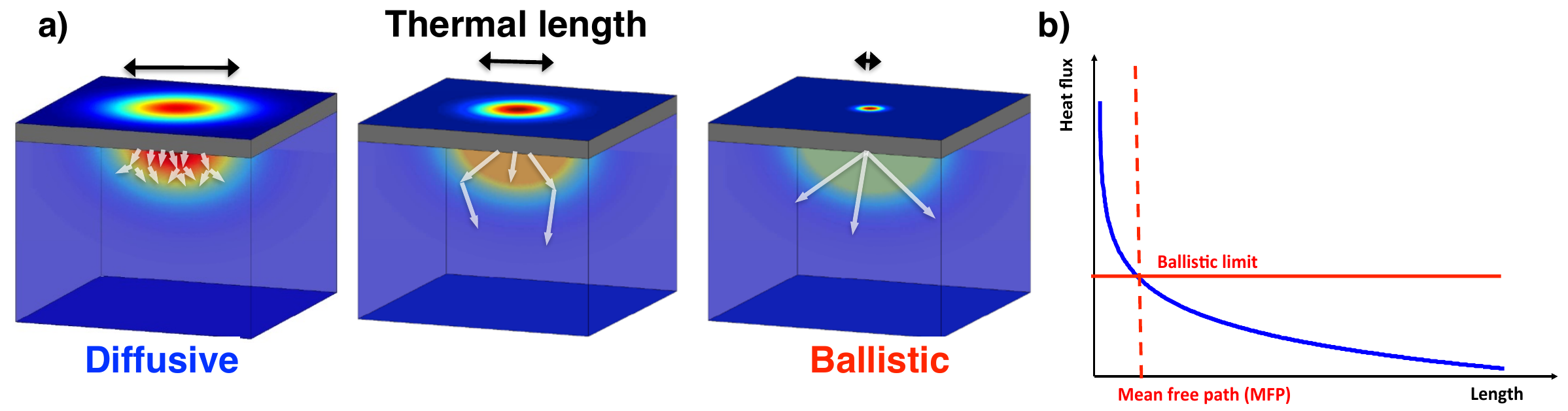}
\caption{(a) Schematic of the MFP spectroscopy technique. The heat flux dissipated by a fixed temperature difference is very different depending on the thermal length relative to phonon MFPs. MFP spectroscopy consists of systematically observing the transition from the diffusive to the ballistic regimes by varying a thermal length scale, from which the underlying MFP distribution can be obtained. (b) Heat flux versus length plot that illustrates the physical principle of MFP spectroscopy. When applied to the ballistic regime, Fourier's law predicts that a finite temperature difference can dissipate a nearly infinite heat flux, an unphysical prediction. In fact, the heat flux cannot exceed the ballistic limit (red horizontal line) in which all phonons propagate at the group velocity without scattering. MFP spectroscopy uses these discrepancies to obtain information about the MFPs in the material. }
\label{MFPSprinciple}
\end{center}
\end{figure}

MFP spectroscopy is based on the physical fact that the heat flux dissipated by a given temperature difference is very different depending on the value of thermal length scale over which a temperature difference exists compared to MFPs \cite{Chen:1996}. If the thermal length is much larger than all MFPs, the heat transport is diffusive and accurately described by Fourier's law. However, if the thermal length is smaller than some MFPs, some phonons do not scatter local to the heated region, violating a fundamental assumption of Fourier's law. In this quasiballistic regime, one can easily show that the actual heat flux is smaller than the diffusive prediction for a fixed temperature difference. As schematically illustrated in Fig.\ \ref{MFPSprinciple}a, MFP spectroscopy consists of observing the discrepancies in heat flux that occur as a thermal length is systematically varied from the diffusive to ballistic regimes, from which the underlying MFP accumulation can be obtained. Practically, the discrepancies in heat flux are observed as a thermal conductivity that appears to vary with a thermal length such as the pump beam size \cite{TCPRL} or modulation frequency \cite{Koh:2007, Regner:2013}.

It is instructive to compare the principle of MFP spectroscopy with previous efforts to infer MFPs by changing the physical dimensions of a sample. For example, Ju and Goodson reported thermal conductivity measurements of Si membranes with variable thickness, from which information about MFPs was obtained by correlating the dimensions of the sample with the thermal conductivity value \cite{Ju:1999}. The key insight of MFP spectroscopy is that one need not change the dimensions of the sample to probe MFPs. Instead, varying the length scale over which heat is transported in a macroscopic material allows the same information to be obtained and without the additional complication of boundary scattering. This insight is essential because it allows the thermal length to be externally controlled and applied to a wide variety of samples with essentially no change in experimental setup.

One subtlety associated with the physical principle underlying MFP spectroscopy is why the ballistic heat flux is smaller than the Fourier law prediction. If scattering events are removed as in ballistic transport, it seems intuitive that the heat flux should be larger than in the diffusive case. Here, one must be very careful about precisely what is being compared. Consider two thermal reservoirs at different, fixed temperatures connected by a crystal with some thickness. When the crystal is made thinner and scattering events become less likely, the heat flux does increase, eventually reaching the ballistic limit in which heat is transported by phonons at the group velocity without scattering. This ballistic limit is the maximum heat flux that can be supported in the material. However, applying Fourier's law based on the thermal conductivity corresponding to the bulk material will predict a heat flux that exceeds this ballistic limit, even tending to infinity as the crystal thickness approaches zero. This diffusion theory prediction is illustrated schematically in Fig.\ \ref{MFPSprinciple}b. Clearly, it is not possible for a finite temperature difference to dissipate a nearly infinite heat flux, and under these conditions the actual ballistic heat flux is smaller than the Fourier law prediction, as stated above. While this example is strictly only valid for steady heat conduction between blackbodies, recent solutions of the BTE by us and others demonstrate that the same physical principle applies in every case, including transient transport, examined thus far \cite{Minnich:2011, Hua:2014, Ding:2014, Maznev:2011b, Minnich:2012, Collins:2013}.

The introduction of MFP spectroscopy is quite recent, although the possibility of using observations of nondiffusive transport to infer phonon relaxation times was raised as early as 1971 \cite{Simons:1971}. More recently, Koh and Cahill observed a modulation-frequency dependent thermal conductivity in semiconductor alloys using TDTR and attributed it to ballistic transport \cite{Koh:2007}. Subsequently, Siemens et al. reported the observation of strongly nondiffusive transport in nickel nanolines patterned on a sapphire substrate, which was attributed to a ballistic thermal boundary resistance \cite{Siemens:2009}. Shortly thereafter, Minnich et al. described the principle of MFP spectroscopy based on observations of a pump-size dependent thermal conductivity in silicon at cryogenic temperatures using TDTR \cite{TCPRL}. Johnson et al. used transient grating spectroscopy to determine MFPs in Si membranes \cite{Johnson:2013}. Regner et al. reported measurements of the MFP distribution in Si and other materials using a new experimental setup called broadband-frequency domain thermoreflectance (BB-FDTR). \cite{Regner:2013} The same authors also reported evidence of a universal MFP distribution in small unit cell semiconductors using the same method \cite{Freedman:2013}.

\begin{figure}
\begin{center}
\includegraphics[width=.9\textwidth]{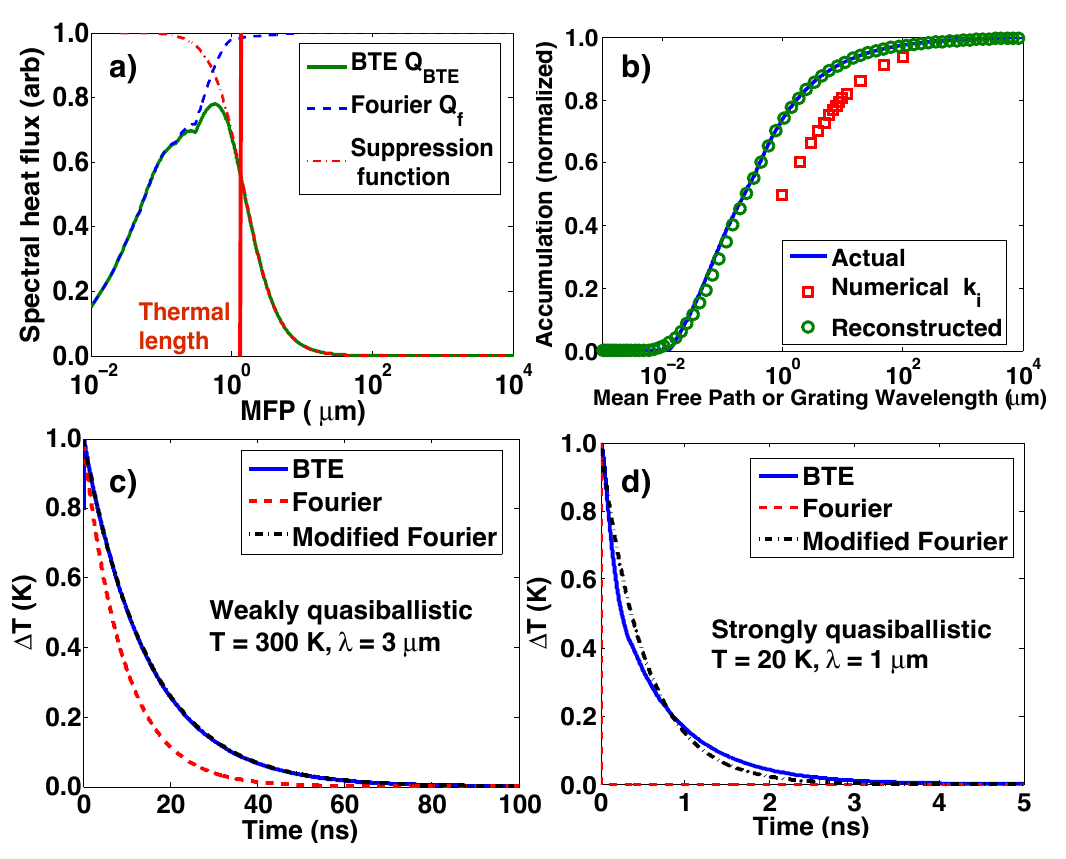}
\caption{(a) Spectral heat flux versus MFP for the transient grating experiment obtained from an analytic solution to the BTE \cite{Hua:2014}, demonstrating that phonons with MFPs comparable to the wavelength of the spatially sinusoidal heating beam (thermal length, red vertical line) have a reduced heat flux (green solid line) compared to the Fourier law prediction (dashed blue line). The magnitude of the suppression is described by a suppression function (red dash dotted line). (b) Numerical example of the reconstruction procedure to obtain the MFP distribution. We used the BTE to simulate several transient grating decays as a function of the spatial wavelength, from which we obtained the effective thermal conductivities (red squares). We then performed the reconstruction procedure to obtain the MFP distribution (green circles), demonstrating excellent agreement with the actual MFP distribution (blue solid line). (c) Temperature decay of a transient grating versus time in the weakly quasiballistic regime, demonstrating good agreement between the BTE result and a modified Fourier's law with an effective thermal conductivity. In this regime, the formal solution of the BTE is a modified solution of the heat equation. (d) Temperature decay of a transient grating versus time in the strongly quasiballistic regime. Here, a modified diffusion theory no longer accurately describes the decay.}
\label{fig:TGsim}
\end{center}
\end{figure}

Despite all of these experimental observations, fundamental questions remain about how to interpret the results. How is the desired quantity, the MFP spectrum, related to the variable thermal conductivities? More generally, what information about phonons is contained in observations of the quasiballistic regime? A number of works,\cite{Cruz:2012}  such as a two-temperature model for ballistic and diffusive phonons, \cite{Wilson:2013} have been recently reported to address these questions. However, while these approaches certainly yield insight, they lack predictive power because they are a simplification of the BTE and thus require assumptions to close the problem.

To avoid this limitation, our approach has been to study quasiballistic transport by rigorously solving the frequency-dependent BTE for which the only input is the phonon dispersion and relaxation times. In particular, we have introduced a theoretical framework that allows the MFP accumulation distribution to be quantitatively reconstructed from thermal measurements \cite{Minnich:2012}. This work is based on a theoretical analysis of the transient grating experiment, which monitors the thermal decay of a sinusoidal initial temperature distribution on a sample. The key observation from this work is shown in Fig.\ \ref{fig:TGsim}a, which plots the calculated spectral heat flux versus the MFP in a transient grating geometry. While Fourier's law predicts that long MFP ballistic phonons contribute substantially to heat conduction, the calculated heat flux from phonons with MFPs comparable to the spatial transient grating wavelength is smaller than this prediction. It is this suppression in heat flux due to quasiballistic transport that is the origin of the thermal conductivities that appear to depend on thermal length scale in experiments \cite{Johnson:2013}. This reduction in heat flux can be described by a suppression function, also shown in the figure, that depends primarily on the experimental geometry. This observation demonstrates that the transition from the diffusive to the ballistic regimes is not a sharp cutoff, as was previously assumed, but occurs over a transition region defined by a thermal length scale.

Based on this observation, we introduced a framework by which the desired MFP distribution can be linked to the measured experimental thermal conductivities by integration with the suppression function that describes the reduction in heat flux that occurs, compared to the Fourier's law prediction, in the geometry of the experiment:

\begin{equation} \label{eq:ki}
k_{i} = \int _{0}^{\infty} S(\lw/L_{i}) f(\lw) d\lw
\end{equation}

where $k_{i}$ are the measured thermal conductivities that depend on thermal length $L_{i}$, $S(\lw/L_{i})$ is the suppression function, $\lw$ is the MFP, and $f(\lw)$ is the desired MFP distribution. This equation is a classic ill-posed inverse problem, but we were able to show that the MFP distribution can still be recovered using a convex optimization procedure. The result of this procedure is shown in Fig.\ \ref{fig:TGsim}b. In this figure, we simulated the temperature decay of a transient grating using the BTE to obtain the effective thermal conductivities, then performed the reconstruction procedure to obtain the MFP distribution using our knowledge of the suppression function. We observed excellent agreement with the MFP distribution used in the BTE simulations, confirming the validity of this approach. 

The attractive feature of this framework is that the BTE need not be solved every time one wants to explain experimental data. Instead, the BTE only needs to be solved once for a particular experimental geometry to extract the suppression function. Thus the key to measuring MFP spectra is to perform thermal measurements over a wide range of length scales and to determine the suppression function for the specific experimental geometry. 

This approach was originally developed for the transient grating experiment, for which the suppression function is known analytically  \cite{Maznev:2011b, Hua:2014, Collins:2013} and confirmed by numerical simulations \cite{Minnich:2012}. Recently, we have been able to provide additional insight into quasiballistic transport in this experiment using a new analytical solution to the frequency-dependent BTE  \cite{Hua:2014} obtained using Fourier transforms \cite{Collins:2013}. An important question we were able to address is why a modified diffusion theory is even valid at all to interpret observations of nondiffusive transport. From our analytic BTE solution, we identified quasiballistic transport regimes that are distinguished by the phonon relaxation times compared to the thermal decay time. In the limit that the thermal decay time is much longer than relaxation times yet with some MFPs longer than the thermal length scale, we found that the exact solution to the BTE is a modified diffusion theory with a reduction in thermal conductivity described by Eq.\ \ref{eq:ki}. Figure \ref{fig:TGsim}c shows the excellent agreement between the actual decay and the decay predicted by a modified diffusion theory. This result provides theoretical justification for the use of a modified diffusion theory to describe nondiffusive transport and is consistent with previous theoretical works \cite{Maznev:2011b, Collins:2013} and experimental observations \cite{Johnson:2013}. When the thermal decay time is comparable to relaxation times, the thermal decay does not follow the functional form predicted by Fourier's law, as in Fig.\ \ref{fig:TGsim}d. However, we found that an effective thermal conductivity can still be defined and used to obtain the MFP spectrum. This result thus formally supports the reconstruction approach of Ref. \onlinecite{Minnich:2012} for the transient grating experiment.

The situation is considerably more complicated for other optical experiments such as time-domain thermoreflectance (TDTR) and broadband frequency-domain thermoreflectance (BB-FDTR). In these experiments, both cross-plane and radial heat conduction play a role, and the presence of a transducer-substrate interface also introduces difficulties in interpretation. At present, a number of puzzles remain, such as the apparent discrepancy between TDTR and BB-FDTR measurements.\cite{Regner:2013, Koh:2007} However, progress is still being made. We are using the same efficient MC algorithms as described in section \ref{sec:MC} to study quasiballistic transport in TDTR \cite{Ding:2014}. We have confirmed that the origin of pump-size dependent thermal conductivities observed previously \cite{TCPRL} are due to quasiballistic transport. Further, we were able to use the spectral information in our simulations to obtain a radial suppression function, enabling MFP measurements with TDTR. At present, we are examining the effect of the interface on the interpretation of the measurements using a new semi-analytical solution of the 1D BTE.

\begin{figure}
\begin{center}
\includegraphics[width=1\textwidth]{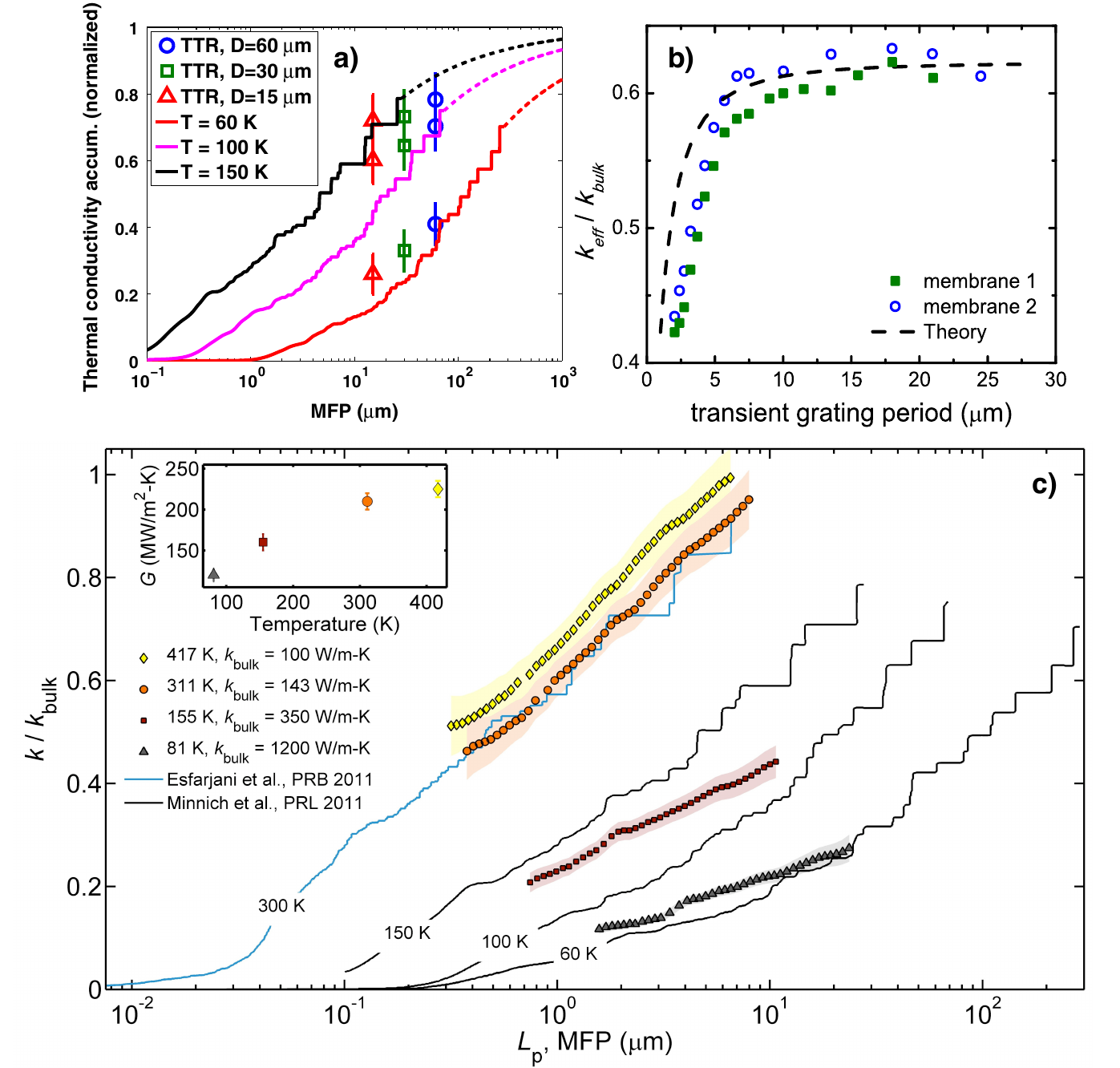}
\caption{Reported measurements using MFP spectroscopy. (a) MFPs in crystalline Si at cryogenic temperatures, showing that phonon MFPs can exceed hundreds of microns due to a lack of phonon-phonon scattering \cite{TCPRL}. (b) MFPs in a 400 nm thick freestanding Si membrane measured using transient grating spectroscopy. Analysis of these measurements \cite{Minnich:2012} indicates that most thermal phonons have MFPs shorter than 1 micron due to diffuse scattering at the membrane boundaries. (c) MFPs in crystalline Si at various temperatures obtained by Regner et al. using BB-FDTR, confirming the theoretical prediction that approximately 40\% of the heat is carried by phonons with MFPs exceeding 1 micron even at room temperature. Reprinted by permission from Macmillan Publishers Ltd: Nature Communications, Ref.\ \onlinecite{Regner:2013}, Copyright (2013).}
\label{MFPSresults}
\end{center}
\end{figure}

Despite these issues, MFP spectroscopy has already provided a number of important insights into heat conduction. A particularly interesting result has been the confirmation of the importance of long MFP phonons to thermal transport in semiconductors despite their small contribution to specific heat. In bulk crystalline Si, MFP spectroscopy has shown that MFPs are hundreds of microns long even at 100 K as shown in Fig.\ \ref{MFPSresults}a \cite{TCPRL}. Measurements performed using transient grating spectroscopy on 400 nm thick freestanding Si membranes are shown in Fig.\ \ref{MFPSresults}b. Analysis of these measurements \cite{Minnich:2012} indicates that most thermal phonons have MFPs shorter than 1 micron due to diffuse scattering at the membrane boundaries. MFP distribution measurements for bulk Si at various temperatures reported by Regner et al. using BB-FDTR are shown in Fig.\ \ref{MFPSresults}c, indicating that phonons with MFPs longer than 1 micron contribute 40\% of the total thermal conductivity at room temperature, in agreement with theoretical calculations \cite{Regner:2013}. As the theory underlying MFP spectroscopy is refined, further novel results are expected to follow.

\section{Summary and Outlook}

A fundamental microscopic understanding of phonon heat conduction has long been desired for both science and engineering but historically difficult to obtain due to limitations on computational and experimental methods. In this topical review, we have described how advances in a diverse set of techniques, most of which were introduced in the past five years, are enabling an unprecedented microscopic picture of thermal phonon transport. The computational techniques include ab initio approaches using density functional theory to describe phonon-phonon interactions, atomistic Green's functions to determine scattering rates from extended defect structures, and variance-reduced Monte Carlo algorithms to simulate mesoscale thermal transport in structures with dimensions as large as tens of microns. Experimentally, advances in inelastic neutron scattering techniques and a new method called mean free path spectroscopy allow direct access to the thermal phonon spectrum for the first time. Collectively, these techniques have provided unique insights into phonon heat conduction, including the importance of considering the phonon MFP spectrum rather than an average MFP, a new view of the microscopic requirements for good thermal conductors, and an improved understanding of phonon transmission across rough interfaces.

What is next for this microscopic investigation? Despite these advances, many aspects of heat conduction in the majority of materials remains unclear. Here, we describe a few of these important questions.

Our understanding of anharmonic scattering has been dramatically improved by ab initio calculations, but many predictions for materials ranging from nitrides to graphene remain to be experimentally verified. An important task will thus be to use INS or MFP spectroscopy to directly measure microscopic quantities such as MFP spectra and compare the results to ab initio calculations. The agreement, or lack of agreement, will lead to a deeper understanding than that obtained from either method separately. An additional interesting topic for anharmonic scattering is understanding phonon interactions in materials with complex unit cells such as Yb$_{14}$MnSb$_{11}$ \cite{Brown:2006}, for which numerous optical branches exist that could scatter acoustic phonons. The unit cells in such materials are so large that they may not be suitable for DFT at present, making an experimental approach the most promising option to gain insight into these materials.

After understanding anharmonic scattering, the next step is to investigate basic scattering mechanisms such as point defect scattering and electron-phonon scattering as they play an important role in thermoelectric materials. The scattering rate due to point defects can be derived from perturbation theory \cite{Tamura:1983} but it has not been experimentally verified. Our understanding of electron-phonon scattering is more elementary, with most treatments being based in Ziman's original treatment from over 50 years ago that suggests that low energy phonons are scattered by electrons \cite{Ziman:2001}. Again, this prediction has never been verified. INS or MFP spectroscopy could be used, although MFP spectroscopy can be more easily applied to a variety of materials.

Subsequently, the important task is to understand scattering due to extended defects such as grain boundaries, superlattice interfaces, nanoparticles, and other natural or artificial structures. Despite the numerous demonstrations of nanostructured materials with reduced thermal conductivities, precisely which phonons are scattered by different defects remains unknown, and even basic questions remain. For example, what is the specularity parameter of phonons incident on a rough boundary? Which parts of the phonon spectrum are scattered most efficiently by which structures? What is the best way to incorporate all of these structures into a material to obtain the minimum thermal conductivity? Conversely, how can heat be most effectively extracted from a volume that includes interfaces, such as the active region of an LED?

The answers to these questions will have an important impact on the science and engineering of thermal conductivity. From a scientific perspective, knowledge of these microscopic details allows us to understand why certain materials are good (or poor) thermal conductors, how specific phonon modes interact with other phonons and defects, and which phonons are responsible for heat conduction. From an engineering perspective, this microscopic information provides the understanding necessary to manipulate the thermal phonon spectrum for numerous applications ranging from thermoelectrics to LED lighting. The thermal transport field thus has an exciting time ahead as new tools are used to answer these interesting and important questions.

\section{Acknowledgements}

The author thanks Sangyeop Lee, Lingping Zheng, Zhiting Tian, Jivtesh Garg, and Olivier Delaire for commenting on the manuscript. This work was sponsored in part by Robert Bosch LLC through Bosch Energy Research Network Grant no. 13.01.CC11, by the National Science Foundation under Grant no. CBET 1254213, and by Boeing under the Boeing-Caltech Strategic Research \& Development Relationship Agreement.

\bibliographystyle{is-unsrt}
\bibliography{ReviewJPCM}

\end{document}